\documentclass[preprint,12pt]{elsarticle}

\def\lbldef#1#2{\expandafter\gdef\csname #1\endcsname {#2}}

\def\href#1#2{#2}

\usepackage{color}
\usepackage{graphicx}
\usepackage{amssymb}
\journal{Physics of the Dark Universe}

\begin{document}

\begin{frontmatter}

\title{The expansion rate of the  intermediate universe in light of Planck}

\author[1,2]{Licia Verde}
\author[3]{Pavlos Protopapas}
\author[1,3]{Raul Jimenez}

\address[1]{ICREA \& ICCUB, University of Barcelona, Barcelona 08028, Spain.}
\address[2]{Institute of theoretical astrophysics, University of Oslo, 0315 Oslo, Norway.}
\address[3]{Institute for Applied Computational Science, Harvard University, MA 02138, USA.}

%\email{liciaverde@icc.ub.edu}
%\email{raul.jimenez@icc.ub.edu}
%\email{pavlos@seas.harvard.edu}

\begin{abstract}
We use cosmology-independent measurements of the expansion history in the redshift range $0.1\lesssim z <1.2$ and compare them with the Cosmic Microwave Background-derived expansion history predictions. The motivation is to investigate if the tension between the local (cosmology independent) Hubble constant $H_0$ value and the Planck-derived $H_0$ is also present at other redshifts. We conclude that there is {\em no tension} between Planck and cosmology independent-measurements of  the Hubble parameter $H(z)$ at $0.1 \lesssim z < 1.2$ for the $\Lambda$CDM model (odds of tension are only 1:15, statistically not significant). Considering extensions of the $\Lambda$CDM model  does not improve these odds (actually makes them worse),  thus favouring the simpler model over its extensions. 
On the other hand the $H(z)$ data are also not in tension with the local $H_0$ measurements but the combination of all three data-sets shows a highly significant tension (odds $\sim 1:400$). Thus the new data  deepen the  mystery of the mismatch between Planck and local $H_0$ measurements, and cannot univocally determine wether it is an effect localised  at a particular redshift. Having said this, we find that  assuming the NGC4258 maser distance as the correct anchor for $H_0$, brings the odds to comfortable values.

Further, using {\em only} the expansion history measurements we constrain, within the $\Lambda$CDM model, $H_0 = 68.5 \pm 3.5$ and $\Omega_m = 0.32 \pm 0.05$ without relying on any CMB prior. We also address the question of how smooth the expansion history of the universe is given the cosmology independent data and conclude that there is no evidence for deviations from smoothness on the expansion history, neither variations with time in the value of the equation of state of dark energy.
\end{abstract}

\end{frontmatter}

\section{Introduction}

The recent  release  of the determination of cosmological parameters   from the Planck \cite{Planck1,Planck2} Cosmic Microwave Background (CMB) observations, has shown that for a $\Lambda$CDM model, the extrapolated value at $z=0$ of the Hubble parameter, $H_0$, is in tension with the one measured locally via astronomical observations of the local distance scale \cite{local,tension}. This could be due to unaccounted systematics in the data in either (or both) experiment, or to a failure of the adopted cosmological model to describe nature.

In a previous paper \cite{local} we discussed the importance of local ($z =0$) measurements in order to assess the consistency of the currently favoured cosmology model: $\Lambda$CDM. Because the CMB mostly probes the universe at $z \approx 1100$, any cosmological parameter defined  at any other redshift is necessarily model-dependent. Therefore, if cosmological parameters can be measured  directly, precisely and robustly at other redshifts than $z \approx 1100$, the comparison with the same quantities derived from the CMB using the standard $\Lambda$CDM model can serve as a test of the model itself.

In Ref.~\cite{tension} we concluded that local measurements of the expansion rate ($H_0$) and of the age of the Universe were in tension (odds 1:53) with the Planck-derived parameters from the CMB within the $\Lambda$CDM model. The tension was driven by $H_0$ and not by the age. With only the data-sets considered there, it was however not possible to determine wether this tension is a signature of systematics in either measurement or new physics: independent data are needed to make further progress.

One way to further investigate this is to ``fill the gap" by  focusing on $z>0$ but still $z\ll 1100$. This can be done by using the recent measurements of $H(z)$ between redshift $0.1 \lesssim z < 1.2$  from the cosmic chronometers project \cite{JL,Simon,stern,moresco,Moresco2012}.  This adds cosmology-independent measurements of the expansion rate back to when the Universe was only $\sim 1/3$ of its current age ($1/3$ to the distance of last-scattering), thus significantly increasing the volume surveyed in the Universe to test the CMB-derived cosmology model. The cosmic chronometer method is the only method that provides cosmology-independent, direct  measurements of the expansion history of the universe. In fact, Supernovae data measure the luminosity distance-redshift relation, which is  related to an integral of $H(z)$. However  the necessary marginalisation over the  (unknown) intrinsic magnitude of the standard candles is effectively equivalent to a marginalisation over an overall amplitude  i.e., over  $H_0$. Baryon acoustic oscillations must rely on the CMB measurement of a standard ruler (the  sound horizon at radiation drag)  to  extract  $H(z)$ information from radial clustering; angular clustering yields a combination of the angular diameter distance and the  sound horizon at radiation drag and angle averaged clustering yields a combination of angular diameter distance, $H(z)$ and sound horizon. While the CMB determination of the sound horizon is robust e.g., Ref.~\cite{Eisenstein/White:2004} it is still somewhat model-dependent (e.g., \cite{Linder/Robbers:2008, debernardis/etal:2009,Mangilli/etal:2010}). Following  Ref.~\cite{Moresco2012} we concentrate on the redshift range $z < 1.2$ where the dependence on the assumed stellar model is small.

In this paper we  explore whether the expansion history at  intermediate redshifts shows any signature of  possible deviations from the CMB-inferred one. The paper is organised as follows: we first investigate (\S \ref{sec:smooth}), in a mostly parameter-independent way via Gaussian Processes (GP), if the $H(z)$ data  show any sign of the expansion not being smooth: we find none. We also find no evidence for variations of the dark energy equation of state parameter (\S\ref{sec:w}). We then explore the constraints on the parameters of several cosmological models (the standard $\Lambda$CDM and its popular extensions) using {\em only} $H(z)$ data and no CMB prior (\S\ref{sec:Hzonly}). Finally we compute the tension between the $H(z)$ data and the Planck-derived expansion history for different cosmological models (\S\ref{sec:tension}).   We conclude in \S\ref{sec:conclusions}.

\section{Is the  expansion history ``smooth"?}
\label{sec:smooth}
Here and in what follows we use the cosmology-independent expansion history measurements from Refs.~\cite{Simon,stern,Moresco2012}.   Ref.~\cite{Moresco2012} showed that at $z>1.2$ the dependence on the assumed stellar population model becomes important. A full treatment that marginalises over the model uncertainty will be presented elsewhere, here we only consider $z<1.2$. In addition we increase slightly (20\%) the error bars of the point at $z=1.04$ to encompass the  stellar model uncertainty. 

We start by  investigating if there are any signatures of deviation from smoothness using Gaussian Processes (GP) \cite{rasmussen2005gaussian, bishop2006pattern}, which is a  non-parameric Bayesian inference formalism.  Ref.~\cite{clarkson} have been the first ones to recently use GP on the same dataset to extrapolate the value of the the Hubble parameter at $z=0$ {\it assuming smoothness} of the  expansion history.  Here we use GP with a different purpose: to determine, in the least parametric way, if the data show any deviation from smoothness as a function of redshift. 

\begin{figure}[h!]
        \centering
                     \includegraphics[width=\columnwidth]{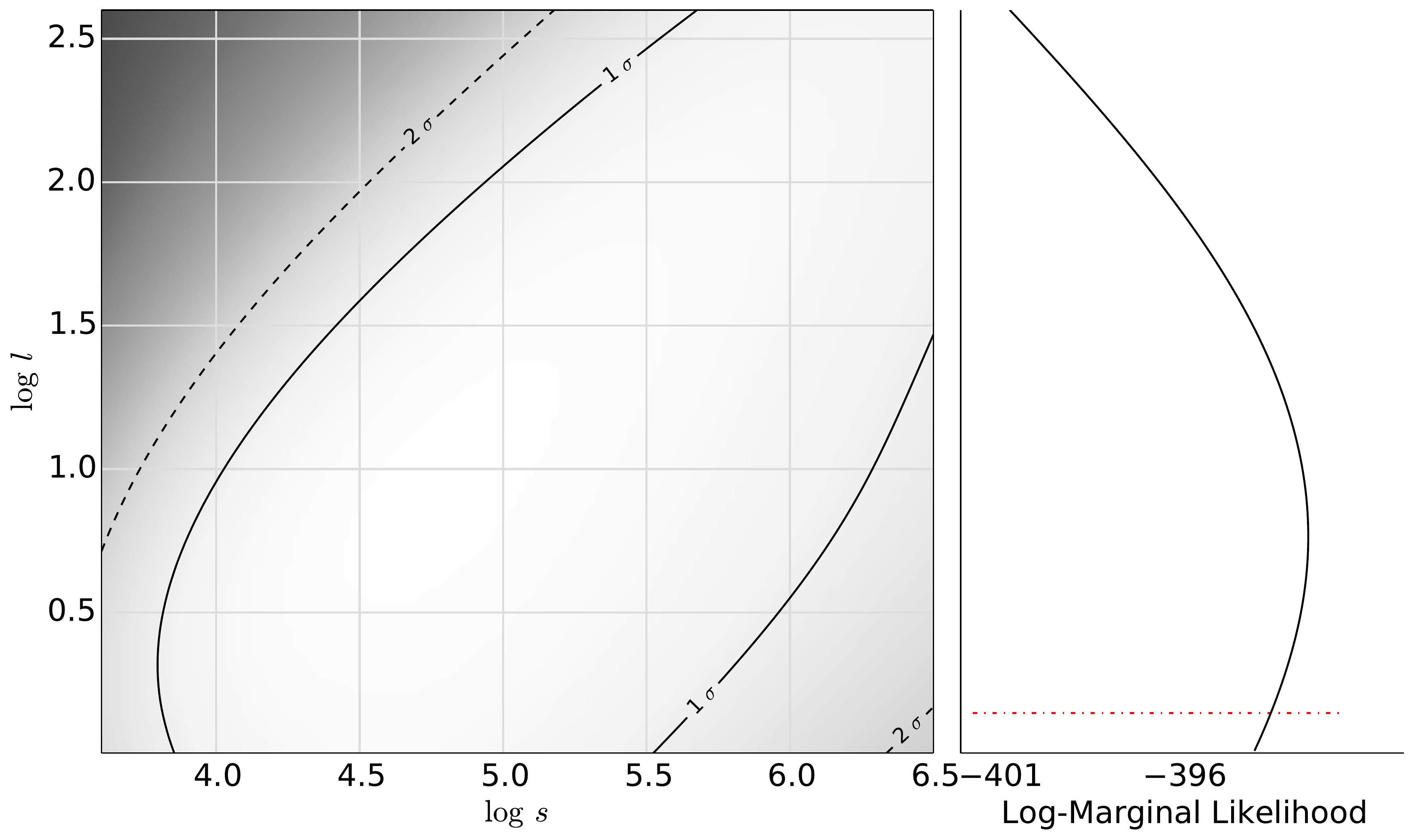}
     \caption{\label{fig:Likelihod} Marginalized (over functions) likelihood given the data as a function of the two hyperparameters, $l$ and $\sigma$. Right panel shows the marginalised (over $\sigma$) of the full likelihood shown on the left. The dotted red line indicates the 5\% confidence of the $l$ hyperparamter. This lower bound gives us the least smooth predictive posterior shown in Fig.~\ref{fig:LCDMwmapplanck}. }
\end{figure}

\begin{figure}[h!]
        \centering
                     \includegraphics[width=\columnwidth,height=2.8in]{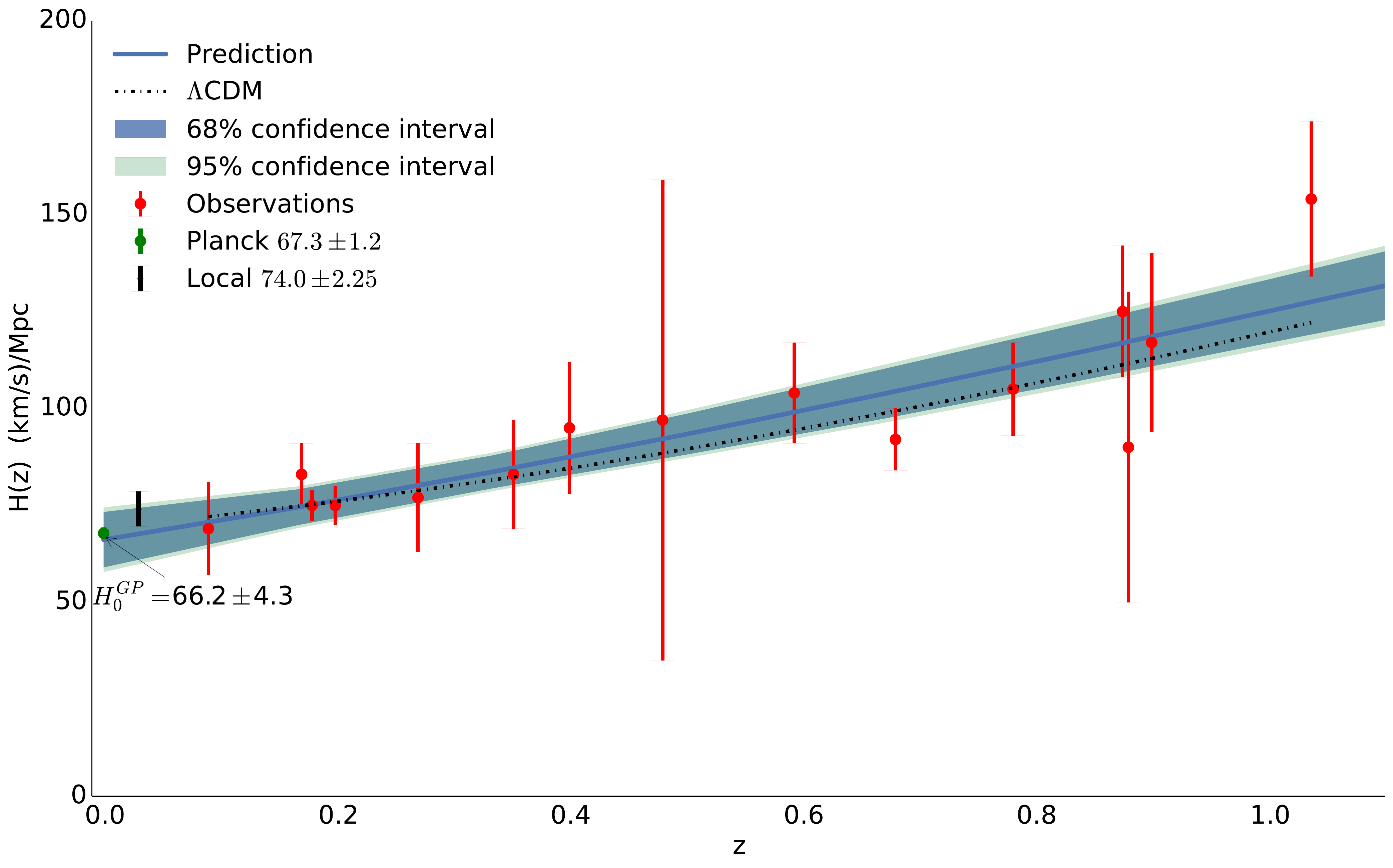}
                     \includegraphics[width=\columnwidth,height=2.8in]{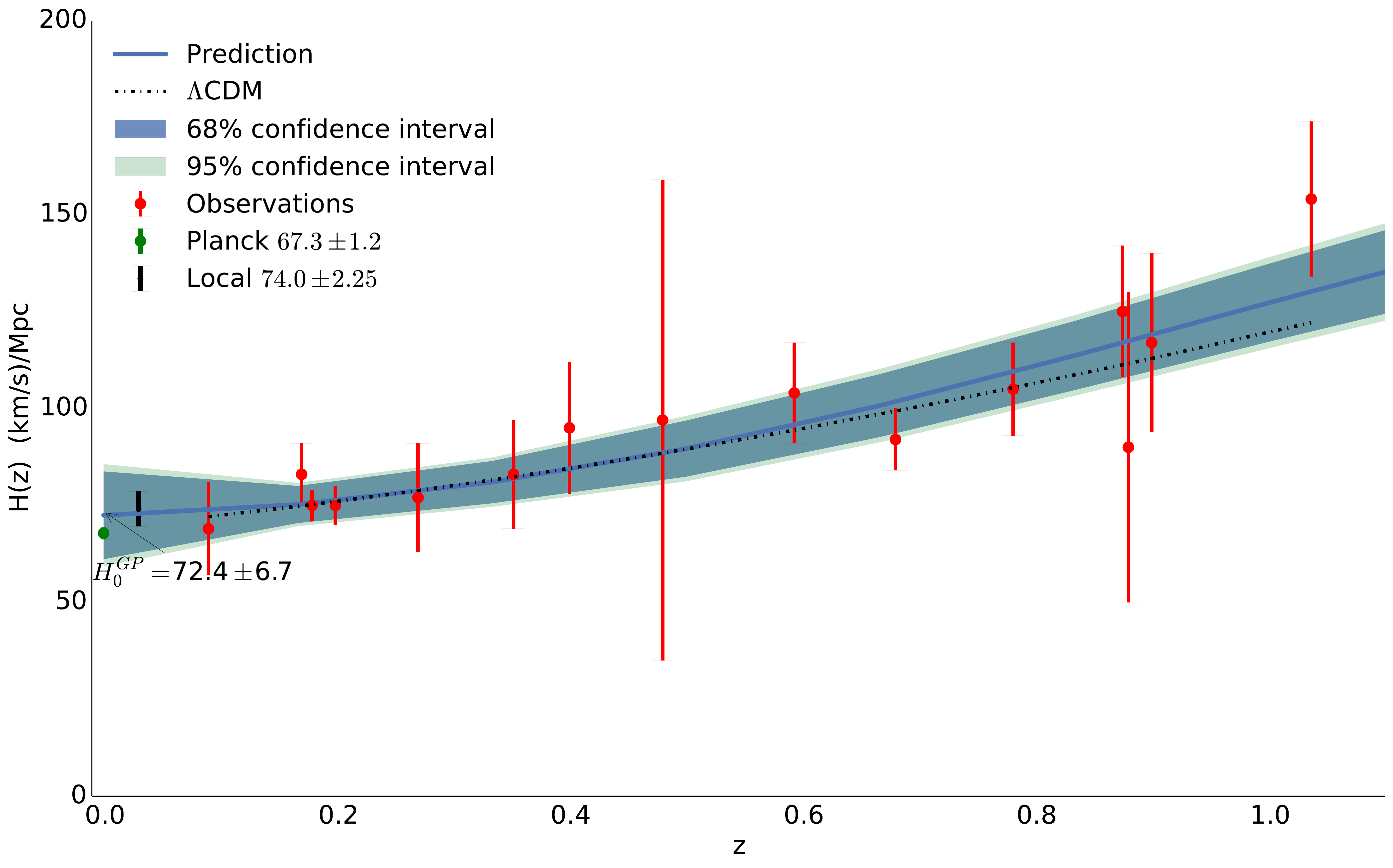}
   \caption{\label{fig:LCDMwmapplanck} Top panel, Gaussian Process prediction (solid regions) of the expansion history of the Universe from the cosmic chronometers data using the best estimate for the hyperparameters. Bottom panel: same prediction but using hyperparameters of low confidence (5\%) that produce the least smooth features (this is achieved with low $l$ values in the covariance matrix).  Note the predicted value for $H_0$. There is no evidence for the expansion history not being smooth over the redshift range $0.1 \lesssim z <1.2$ even at the extreme case of using vey unlikely values of the hyperparameters. For reference the dash-dotted line is the (multi-dimensional) best fit from Planck data. The 68\% confidence error around this line is $\pm 1$ at $z=1.1$. }
\end{figure}

We will not enter into a detailed description of GPs here but instead refer the interested reader to the literature  (see e.g., \cite{rasmussen2005gaussian, bishop2006pattern}  for standard references in the field). GP is a stochastic process that  considers independent and identically distributed Gaussian measurement errors, the  relations or correlations between measurements are defined via a covariance matrix. Using  standard Bayesian methodology, the estimation of $H$ at any value of $z$ is determined by a probability distribution (posterior predictive). This results into a fully probabilistic estimation of the expansion history given the measurements. 
In our application the different $H(z)$ determinations are uncorrelated, therefore the  off-diagonal terms of the covariance matrix determine the smoothness of the fitting function. 
In this approach, the only choice we have to make is the form of the off diagonal terms of the covariance matrix. We have experimented with various  choices of the covariance matrix with similar results, but we report here the outcome of employing the most commonly used covariance matrix, the radial exponential:

\begin{equation}
   K(z,z') = \sigma^2 \exp{ \left( -\frac{(z-z')^2}{2l^2} \right) },
\end{equation} 

\noindent where $z$ and $z'$ are two different z values  (i.e., $z \ne z'$) and $l$, $s$ are the latent parameters or hyper-parameters which are determined using  $H(z)$ data themselves.  Intuitively, $l$ controls the  correlation length and therefore the ``smoothness" and $\sigma$ controls the  importance of the off-diagonal terms compared to the diagonal ones.  
The bigger the $l$ is, the smoother the predictive function would look. Similarly, the higher the $\sigma$ is the lower the signal-to-noise ratio is. The values of $l$ and $\sigma$ can be obtained using maximum likelihood estimation.
Since we are dealing with hyper-parameters and not parameters, we maximise the marginal likelihood (over functions) and not the likelihood directly.  Note that this  approach  is the same as the hierarchical Bayesian  one \cite{MacKay1992}. 
 To do so we  calculate the partial derivative of the marginal likelihood  with respect to  the hyper-parameters and optimise the
hyper-parameters using gradient based search \citep{rasmussen2005gaussian}. 

Fig.\ref{fig:Likelihod} shows the marginalized  likelihood given the data as a function of the two hyperparameters, $l$ and $\sigma$. The right panel shows the marginalized full likelihood (left panel) over $\sigma$. The dotted red line indicates the 5\% confidence of the $l$ hyperparameter. This lower bound gives us the least smooth predictive posterior shown in Fig.~\ref{fig:LCDMwmapplanck}.

The top panel of Fig.~\ref{fig:LCDMwmapplanck}  shows  the 68\% and 95\% confidence bound of the predicted posterior using the GP  with the radial exponential covariance  and the best estimate for the hyper-parameters,  as described above. The bottom panel shows the same prediction but using hyper-parameters that produce the 5\% least smooth features. 

These regions are fully compatible with the Planck CMB derived expansion rate (dot-dashed line). But more interestingly is the fact that there is no signature for deviations from smoothness even when we considered the least likely value of $l$ that creates the least smooth curve. Any sharp or asymmetric  variation of the equation of state parameter, $w$, that results in a non-smooth expansion history,  does not seem to be present at the accuracy of the data. 

Another interesting feature is the value of the extrapolation at $z=0$ (as already pointed out and studied by Ref.~\cite{clarkson}). While the error at $z=0$ is larger than the one from Planck \footnote{The Planck determination reported here assumes a flat $\Lambda$CDM model.}, it is fully compatible with it. It however points to the central value of the locally determined $H_0$ being slightly high. Here we use the Ref.~\cite{local} ``world average" value for $H_0$ obtained from the \cite{Riess/etal:2011} and \cite{Freedman/etal:2012} determinations.

\section{Does w vary with time?}
\label{sec:w}
In this section, we examine possible variations of the dark energy equation of state parameter $w$ as a function of $z$. We use the  $\Lambda$CDM model as the true model and we simply ask if for different domains of $z$ the best estimates of $w$ are statistically different or similar\footnote{We use weakly informative priors: uniform prior on $\Omega_m$ between 0 and 1, Gaussian prior on $H_0$ centred around $68$ Km s$^{-1}$Mpc$^{-1}$ with a width of 10 Km s$^{-1}$Mpc$^{-1}$ and uniform prior on $w$ between -2 and 0.}. We first estimate all three parameters for the whole dataset and then we divide the data into subsets and compare the posteriors of $w$ looking for significant statistical differences. For  easy reference, in the appendix we report useful formulae for the expression of $H(z)$ as a function of cosmological parameters.

\begin{figure}[h!]
        \centering
                     \includegraphics[width=\columnwidth]{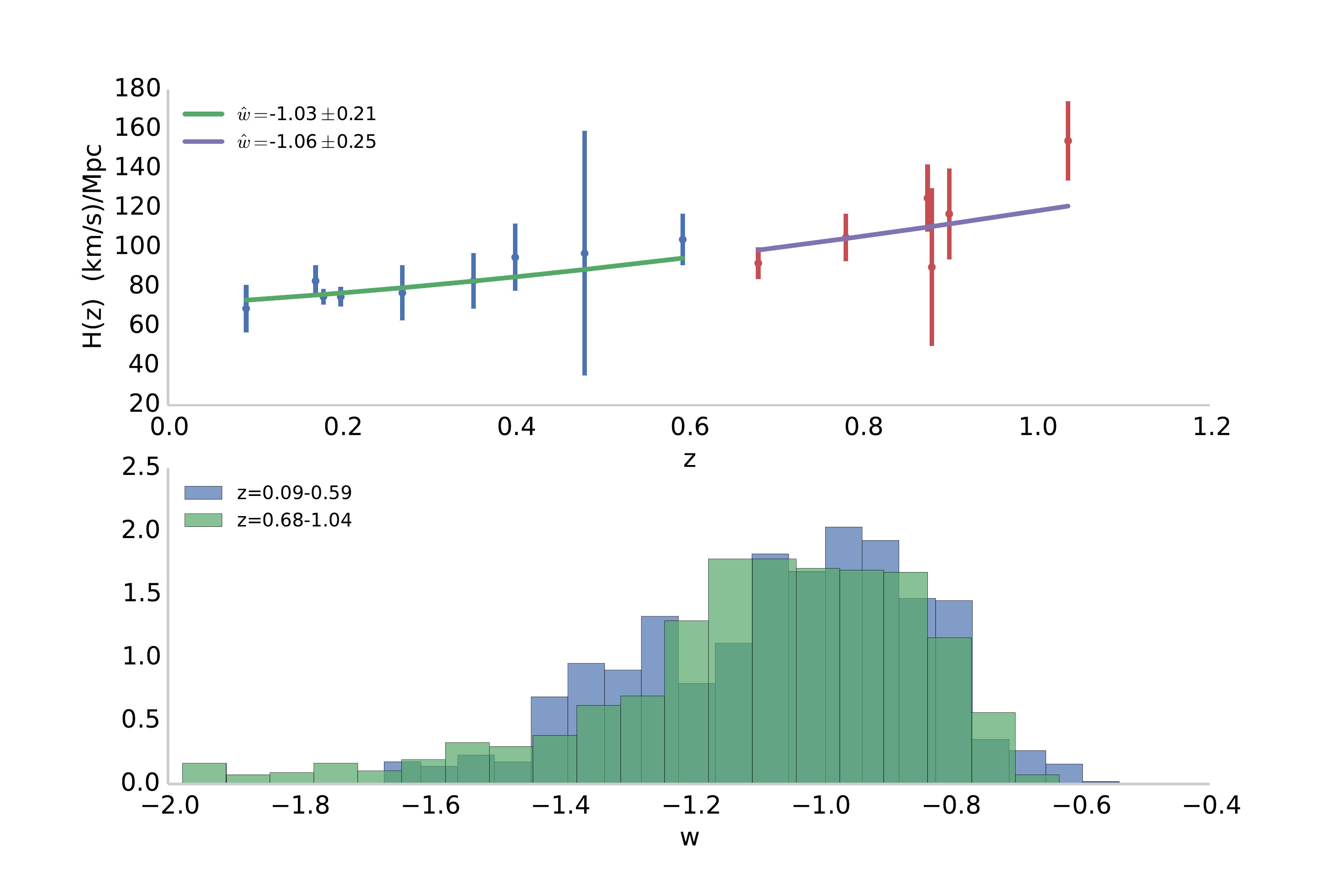}
   \caption{\label{fig:varyingw} Top panel,  mean value of the posterior distribution of $H$ as a function of $z$. In this test, data were divided into two different subsets: $0.1\leq z<0.6$ and $0.6<z<1.2$. Both tests were treated independently but we used the same priors.  Bottom panel shows the posterior for the $w$ for the two data sets. Clearly the two estimates are consistent.  }
\end{figure}

We perform multiple tests by dividing the data into different subsets and repeat the experiment many times. None of these tests produce any significant evidence that $w$ varies above the noise of the measurement given the $\Lambda$CDM model. 
In order to enhance the power of our test, we fixed the values of $H_0$ and $\Omega_m$ to the values estimated from the whole data set and repeat the test looking for statistically different values of $w$, an indication of a varying $w$. This is shown in 
Fig. \ref{fig:varyingw}.  In this particular test the data were divided into two sub-sets and we find the marginalised posteriors for $w$ for both sub-sets. On the top panel we show the results of the fit using Bayesian linear regression, where on the bottom we compared the marginalized posteriors. All test we performed show similar results. We conclude that given the measurement error $w$ is invariant in time. 

\section{What does the expansion history tell us about the cosmological model?}
\label{sec:Hzonly}
\begin{figure}[h]
        \centering
                     \includegraphics[width=.45\columnwidth]{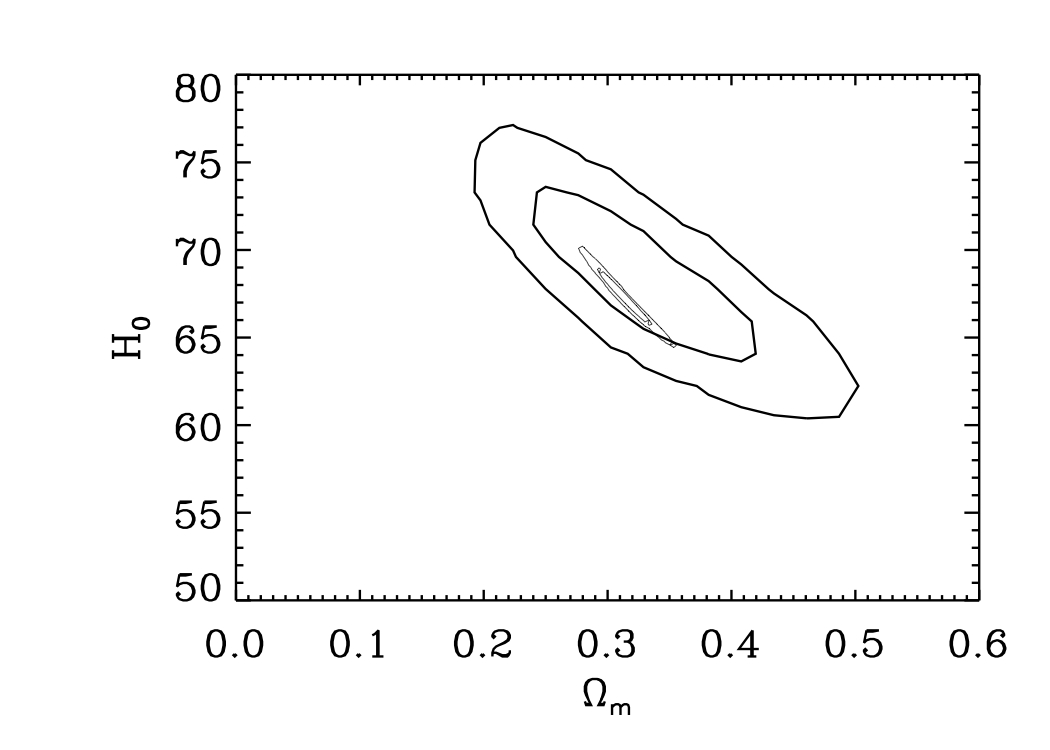}
                     \includegraphics[width=.45\columnwidth]{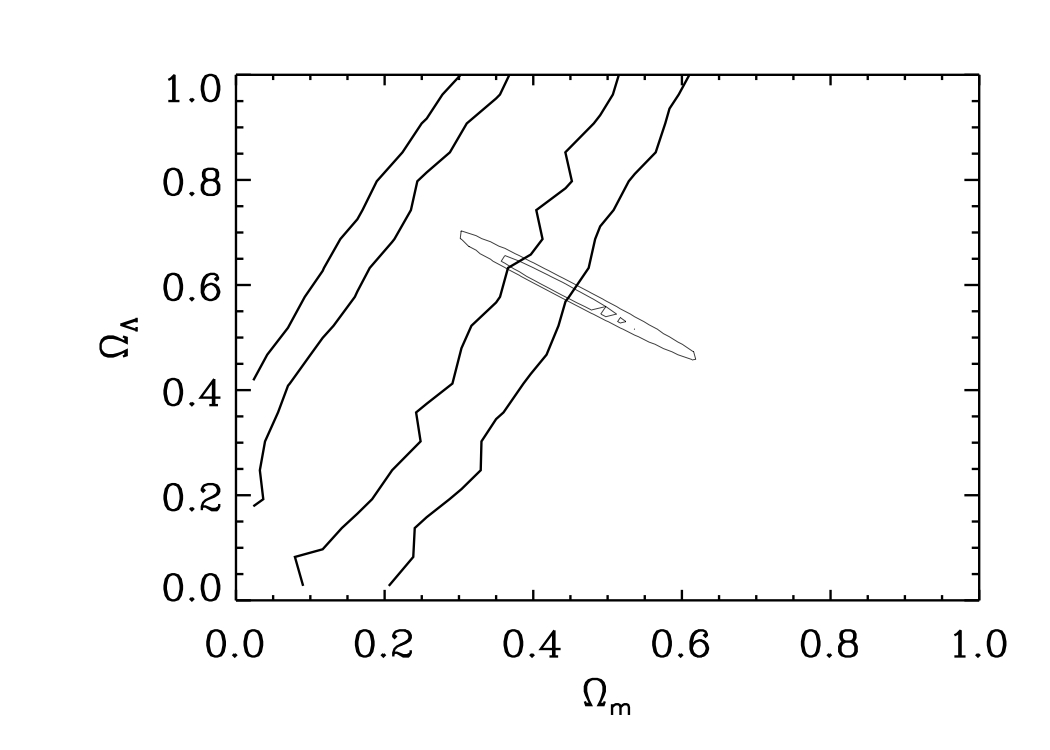}
                                         \includegraphics[width=.45\columnwidth]{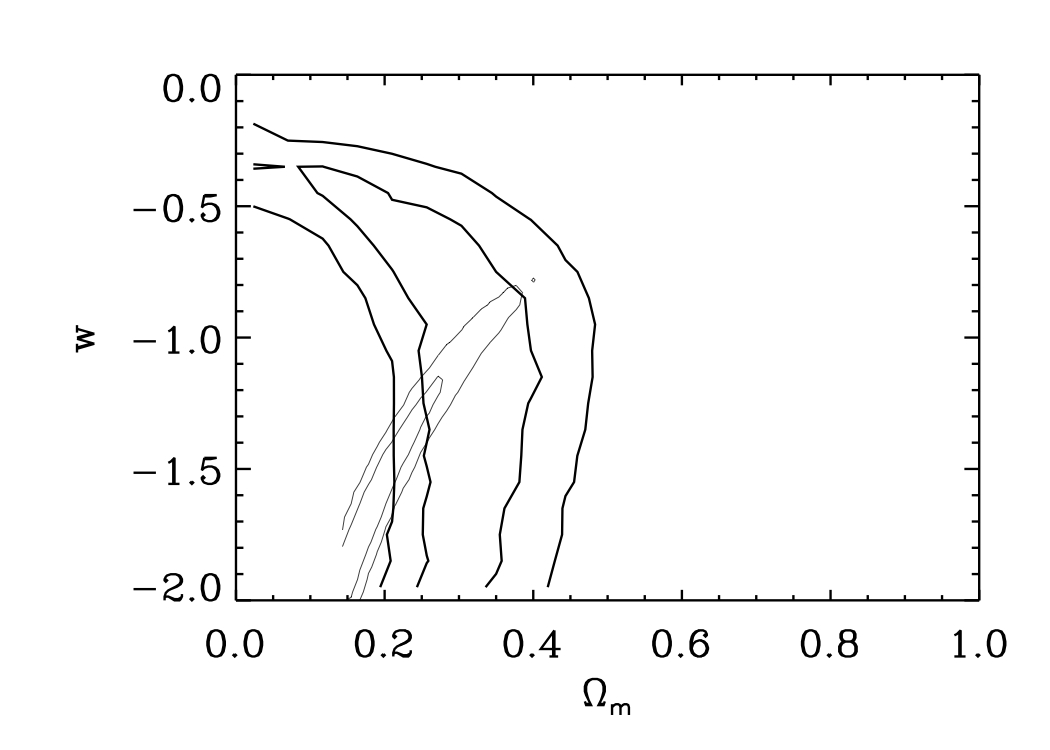}
\caption{\label{fig:lcdm}  68 and 95\% joint confidence contours for different cosmological models. Top left panel: $\Lambda$CDM, top right panel: o$\Lambda$CDM and bottom panel: $w\Lambda$CDM. The thick lines correspond to constraints obtained with expansion history data {\em only}, while the thin contours are using Planck temperature data {\em only}. Note the nice orthogonality and complementarity of both datasets.}
\end{figure}

It is interesting to investigate what constraints on cosmology (cosmological models and parameters)  one can impose by using {\em only} the cosmic chronometers data.
 This is shown in table~\ref{table:params} and Fig.~\ref{fig:lcdm}. The models we consider are: the standard $\Lambda$CDM, $w\Lambda$CDM where the (constant) parameter describing the equation of state of dark energy can differ from $-1$,  o$\Lambda$CDM where we allow the curvature to be non-zero and finally $w_0w_a$CDM, where the Universe is spatially flat  but   the equation of state of dark energy can change as a function of time. The constraints are qualitatively similar (and in agreement with) to the ones obtained from Supernovae type 1 A data, perhaps not surprisingly given that cosmic chronometers and supernovae are probes of the expansion history.  As illustrated in Fig.~\ref{fig:lcdm} the constraints and degeneracy direction in parameter space,  for models that are generalisations of the $\Lambda$CDM are complementary to the ones obtained from the CMB. 

\begin{table}[h]
\begin{center}
\begin{tabular}{ccccccc}
\hline
model & $H_0$ & $\Omega_m$ & $\Omega_{\Lambda}$ & $\Omega_k$ & $w_0$ & $w_a$ \\
\hline
$\Lambda$CDM &  $68.5 \pm 3.5$ & $0.32 \pm 0.05$ & $1 - \Omega_m$ & 0 & $-1$ & $0$ \\
w$\Lambda$CDM & $70 \pm 7$ & $< 0.4$ (95\%) & $1 - \Omega_m$ & 0 & $ < -0.3$ (95\%) & $0$ \\
oCDM & $ > 50$ (95 \%) & -- & -- & -- & $ -1$ & $0$ \\
$w_0w_aCDM$ & $ > 45$ (95 \%) & -- & -- & -- & -- & -- \\
\hline
\end{tabular}
\end{center}
\caption{\label{table:params} Parameter constraints on different cosmology models using the cosmolgy-independent expansion history determined by the cosmic chronometers compilation of  Ref.~\cite{Moresco2012}. When no error-bars are reported the parameter is kept fixed at the reported value. The "-" symbol denotes no meaningful constraint.}
\end{table}

When considering marginalised  constraints on single parameters, table~\ref{table:params} shows that the $\Lambda$CDM model is fairly well constrained for the two parameters that affect the expansion history ($H_0$ and  the density parameter $\Omega_m$). Once the $\Lambda$CDM is extended, the cosmic chronometers data by themselves are not very constraining, being able to provide only upper(lower) limits on the parameters of the cosmology model. It is worth noting that for the most general extension of the $\Lambda$CDM model ($w_0w_a$CDM) $H_0$ has to be larger than $45$ km s$^{-1}$ Mpc$^{-1}$.

\section{Comparing  with the universe at $z=1100$} 
\label{sec:tension}
In Ref.~\cite{tension} we introduced a new statistical tool to determine the odds of two distributions being in tension and thus decide if one could combine the two distributions to further restrict the parameter space of the model. We briefly review here our main argument for completeness and to clarify notation.

Imagine we have performed two experiments: $A,B$ and for each experiment
we produce a posterior $P_{A,B}(\theta|D_{A,B})$ where $\theta$ represents the parameters
of the model and $D_{A,B}$ represents the data from experiments $A,B$ respectively.  Let us also assume that for producing both posteriors we have used the same, uniform priors over the same ``support", x, i.e., $\pi_A=\pi_B=\pi$, $\pi=1$ or $0$ and  therefore $\pi_A \pi_B=\pi$. Let $H_1$ be the (null) hypothesis that both experiments measure the same quantity, 
the models are correct and there are no unaccountable errors.  In this case, the two experiments will produce two posteriors,  which, although can have different (co)variances, and different distributions, have means that  are in agreement.
The alternative hypothesis, $H_{\neg 1}$ is when the two experiments, for some unknown reason,   do not agree, either because of systematic errors or because  they are effectively measuring different things or the model (parameterization) is incorrect. In this case, the two experiments
will produce two posteriors with two  different means and  different
variances. To distinguish the two hypothesis we use the  Bayes factor,
imagine that we can perform a  {\it translation} (shift) of  (one or both of) the distributions in $x$ and let us define $\bar{P}_A$ the shifted distribution. This translation changes the location of the maximum but does not change the shape or the width of the distribution. 
If the maxima of the two distributions coincide then 
\begin{equation}
\int \bar{P}_A \bar{P}_B dx=\bar{\cal E}|_{{\rm max}A={\rm max B}}\,.
\end{equation}
This can be considered our ``straw man" null hypothesis ($H_1$). 
 As the distance between the maxima increases (but the shape of the distributions remains the same),
\begin{equation}
\int \bar{P}_A \bar{P}_B dx=e<\bar{\cal E}\,,
\end{equation}
and eventually $e\longrightarrow 0$ as the two distributions diverge. Clearly the Evidence ratio for the (null) hypothesis $E_1$ is ${\cal E}/\bar{\cal E}|_{{\rm max}A={\rm max B}}$, as the normalization factors $\lambda$ cancel out, and  the Evidence ratio for the alternative $H_{\neg 1}$ is its reciprocal.
We therefore introduce:
\begin{equation}
{\cal T}=\frac{\bar{\cal E}|_{{\rm max}A={\rm max B}}}{\cal E} \, ,
\label{eq:tension}
\end{equation}
  which denotes the degree of tension that can be interpreted in the  widely used (slightly modified, \cite{kassraftery}) Jeffrey's \cite{jeffreys} scale. ${\cal T}$ indicates the odds:   $1\,:\,{\cal T}$  are the chances  for the null hypothesis. In other words, a large tension means that the null hypothesis (${\rm max} A$ = ${\rm max} B$)  is  unlikely.

Continuing on with  the logic outlined above,  let us consider that experiment A gives  us the prior and experiment B gives us the data.  ${\cal T}$ is the evidence  ratio between the integrated {\it posterior} in  two cases. On the numerator: the prior from experiment A has a maximum that coincides with the maximum likelihood from experiment B; on the denominator: the prior from experiment A has a maximum that happened to be where it is.\footnote{Note that here one could loosely exchange ``likelihood'' with ``posterior" for experiment B since we assume uniform priors for experiments A and B that are either 1 or 0 over the same ``support" and therefore one can absorb the prior for experiment B into the A prior.}
Note that since the evidence is translation-invariant it does not matter which data set is considered as prior and which one as ``data": the result is symmetric.

This argument can be extended to as many measurements as we want: the relevant integral just becomes higher-dimensional. For example let us consider the 15 $H(z)$ measurements  in light of a Planck CMB prior for $H(z)$. 

\begin{figure}[h]
        \centering
                     \includegraphics[width=\columnwidth]{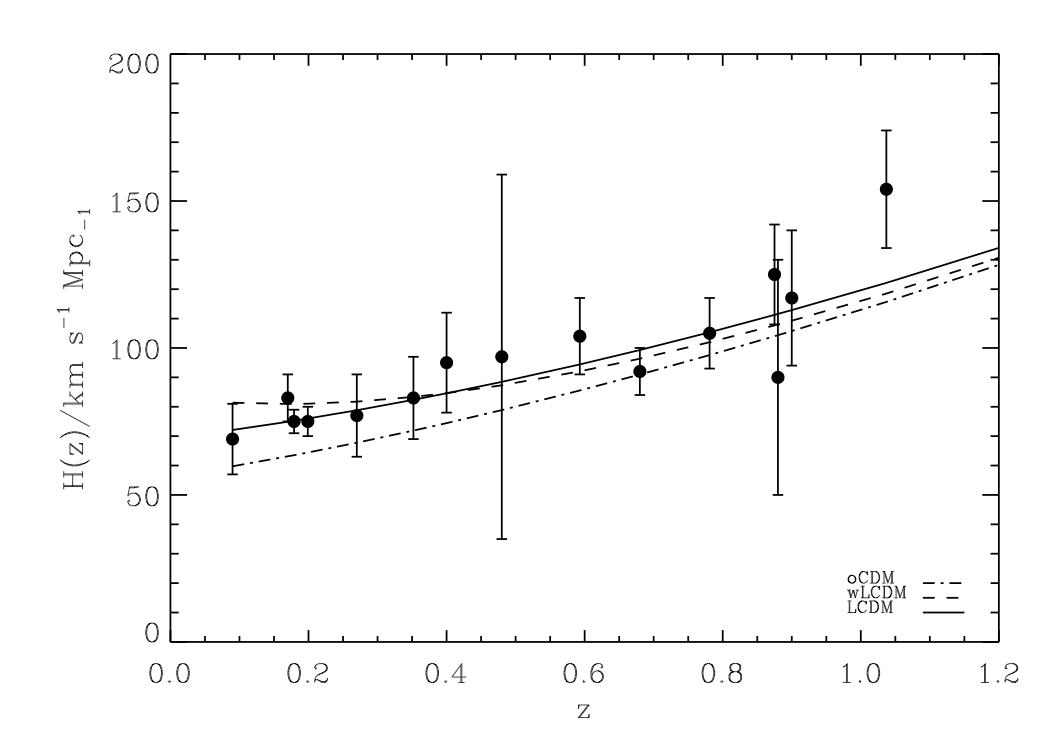}
   \caption{\label{fig:tension} Expansion history from the cosmology-independent cosmic chronometers project compared to different best-fit Planck cosmology models in the redshift range $0.1 \lesssim z < 1.2$.}
\end{figure}

Table~\ref{table:tension} shows the tension and odds for three different cosmology models while Fig.~\ref{fig:tension} shows the corresponding expansion history. 

The first result to note is that the odds of the expansion history data and the Planck CMB $\Lambda$CDM model derived parameters being in tension are very low: just 1:15, which is above 1 $\sigma$ but below 2 $\sigma$; we conclude that for the redshift range $ 1.2 > z \gtrsim 0.1$ there is no evidence for tension between ``local" data and CMB model derived expansion parameters. We also note that extensions to the $\Lambda$CDM model are in tension with the ``local" data, thus disfavouring extensions to the minimal $\Lambda$CDM model.

\begin{table}[ht]
\begin{center}
\begin{tabular}{ccc}
\hline
 model & $ln {\cal T}$ & odds \\
 \hline
$ \Lambda$CDM & 2.73 & 1:15 \\
 $w\Lambda$CDM & 3.27 & 1:26 \\
 o$\Lambda$CDM & 4.93 & 1:138 \\
 \hline
 \end{tabular}
 \caption{\label{table:tension}  Tension (in the Jeffrey's scale and odds that the expansion history data from the cosmic chronometers ($1.2 < z < 0$) and the Planck CMB derived models are in tension. Note that there is no significant tension with the Planck $\Lambda$CDM model. Extensions to this model increase the tension with the model-independent expansion history $H(z)$ data; thus no extension of the model is justified.}
 \end{center}
 \end{table}

Both the Planck CMB  results  and the local $H_0$ ones have been recently re-examined \citep{Efstathiou:2013, Spergel/etal:2014}.

Ref.~\cite{Spergel/etal:2014} excludes the 217 GHz channel from the analysis with the motivation that it  has been shown to fail a number of null tests and therefore could be affected by some systematics. They also  independently re-do the cleaning and analysis. As a result some of the parameters show some shifts, most notably $\Omega_m$ (down by $1\sigma$) and $H_0$ (up by $0.6\sigma$):  $\Omega_m=0.302 \pm 0.015$;  $H_0=68.0 \pm 1.1$. We do not have access to their full posterior therefore we shift the Planck posteriors to  be centred to these values and we refer to this as Planck ``revisited". This  underestimates the final error-bars  by some 20-30\% but it is a conservative estimate of the agreement of the new analysis with lower redshift measurements.
Ref.~\cite{Efstathiou:2013} studies the dependence of the $H_0$ measurement on the choice of distance scale anchor and on the procedure for outliers rejection. LMC  and MW cepheids  distance  anchor yield  values of $H_0$ in agreement with each other but   NGC 4258 maser distance gives lower values. 

Using all distance anchors he obtains
$H_0^{\rm Efs, all}=72.5 \pm 2.5 $km s$^{-1}$Mpc$^{-1}$ while using the maser distance yields  $H_0^{\rm Efs, NGC4258}= 70.6 \pm 3.3 $km s$^{-1}$Mpc$^{-1}$. The combination  LMC + MW yields a value very close to the ``world average" ($H_0^{\rm WA}$ in this table) used above (we do not consider this combination below as results are virtually unchanged).
%
% NGC 4258 maser distance is H0 = 70.6±3.3 km s?1Mpc?1,
%H0= 71.8 ± 2.6 km s?1Mpc?1,  NGC 4258 + LMC
%H0= 72.2 ± 2.8 km s?1Mpc?1, NGC 4258 + MW,
% H0= 73.9 ± 2.7 km s?1Mpc?1, LMC + MW,
% H0= 72.5 ± 2.5 km s?1Mpc?1, NGC 4258 + LMC + MW.
In table \ref{tab:tensionfull} we report the results of the various combinations; the tension between CMB and  cosmic chronometers $H(z)$ remains virtually unchanged between the original Planck and the revisited Planck constraints.

\begin{table}[ht]
\begin{center}
\begin{tabular}{|c|cc|cc|}
\hline
 data combination&\multicolumn{2}{|c|}{Planck orig.}  & \multicolumn{2}{|c|}{Planck Revisited}\\
 & $ln {\cal T}$ & odds &$ln {\cal T}$ &  odds \\                              
\hline
Planck+$ H(z)$ &2.73 &1:15 & 2.71 &1:14\\
 Planck + $H_0^{\rm WA}$ & 3.55 & 1:35 & 2.85 & 1:17 \\
  Planck + $H_0^{\rm Efs, NGC4258}$ &0.44  &1:1.6  & 0.27& 1:1.3 \\
   Planck + $H_0^{\rm Efs, all}$ & 1.77 & 1:6 & 1.33 & 1:3.7 \\
Planck+$H_0^{\rm WA}$+$H(z)$ & 6.17& 1:473 & 5.5 & 1:245  \\
Planck+$H_0^{\rm Efs, NGC4258}$+$H(z)$  & 3.1 & 1:22 & 3.0 & 1:20 \\
Planck+$H_0^{\rm Efs,all}$+$H(z)$  & 4.4 & 1:81 & 4.0 & 1:55 \\
 \hline
 \end{tabular}
 \caption{\label{tab:tensionfull} Tension and odds within the $\Lambda$CDM model,  for different combinations of the datasets considered: Planck orig. refers to the Planck team's analysis, Planck revisited refers to Ref.~\cite{Spergel/etal:2014}, $H_0^{\rm WA}$  refers to the ``word average" of Refs.\cite{Riess/etal:2011, Freedman/etal:2012}  following \cite{local}. $H_0^{\rm Efs, all}$ and $H_0^{\rm Efs, NGC4258}$ are from Ref.\cite{Efstathiou:2013}. Clearly both re-analysis lower the tension.  Note that the numbers for the combination Planck + $H_0^{\rm WA}$ are different from Ref.~\cite{tension} because here we do not include the measurement of the age of the Universe.}
 \end{center}
 \end{table}
 
\begin{figure}[h!]
        \centering
                     \includegraphics[width=0.8\columnwidth]{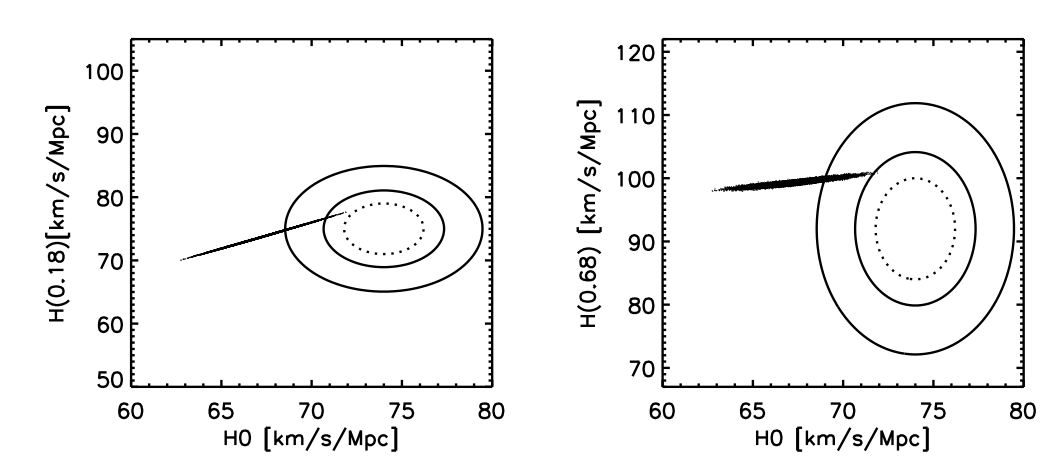}
   \caption{\label{fig:hzmulti} Two-dimensional projection of the posteriors considered. On the left panel $H_0$ vs $H(z=0.17)$ and in the right panel  $H_0$ vs $H(z=0.68)$. The points represent the Planck extrapolated values within the the $\Lambda$CDM model  sampled via Markov Chain Monte Carlo, the ellipses are the direct determinations. }
\end{figure}

 Clearly both re-analyses lower the tension. If only Planck and the local $H_0$ measurements are considered either reanalysis  brings the tension to comfortable values, but if all measurements are considered (i.e.  we consider a  16 dimensional tension) both re-analyses are needed and only the maser distance brings the odds to comfortable values. 
 Note that $H(z)$ +  $H_0^{\rm WA}$  are consistent with each other (odds 1:2.4, $\ln{\cal T}=0.9$). What we see here is the power of  combining many different determinations and therefore evaluating the tension in high-dimensions. When  degeneracies are present (as for example in the case of the Planck extrapolated values of $H(z), H_0$)  single determinations might be in agreement but the multi-dimensional distribution might be less so; this is illustrated in Fig.~\ref{fig:hzmulti}.
 
\section{Discussion and Conclusions}
\label{sec:conclusions}
Using the cosmology independent expansion history data ($H(z)$) from  the compilation of Ref.~\cite{Moresco2012} we have explored if the intermediate redshift universe is in tension with the Planck $\Lambda$CDM derived one. To do so we have generalised the two-dimensional tension introduced in Ref.~\cite{tension} to an arbitrary number of dimensions.  The inclusion of new data adds cosmology-independent measurements of the expansion rate back to when the Universe was only $\sim 1/3$ of its current age ($1/3$ to the distance of last-scattering), thus significantly increasing the volume surveyed to test the CMB-derived cosmology model. We find no tension (odds only 1:15) between Planck and $H(z)$. This is in contrast with the local universe ($z=0$) which is  in  significant tension \cite{tension} (1:57) with Planck. However  $H(z)$ and   $H_0$ are not in tension within a $\Lambda$CDM model: the central value for the Hubble constant obtained from $H(z)$  happens to be very near the Planck-CMB inferred one, but the error-bars are larger. When considering the three datasets--Planck, $H(z)$ and $H_0$-- the  tension becomes highly significant (odds become uncomfortably low  $\sim 1:400$). Thus the addition of $H(z)$ deepens the mystery of the mis-match between Planck and local  $H_0$ measurements, and cannot univocally determine wether it is  an effect localised at $z\sim 0$ or $z \sim 1100$ or if it  shows as a smooth change of the measured value of $H_0$ as a function of the redshift of the measurement.

Both the Planck and the local $H_0$ measurements have been independently re-visited and re-analysed.  Either re-analysis brings the tension between   Planck and local  $H_0$ measurements to comfortable level. On the other hand,  when the three datasets are considered the Planck re-analysis has very little effect, and only assuming the NGC4258  maser distance as the correct anchor for $H_0$ brings the odds to comfortable values.  However at present there are no compelling reasons to  discard the other anchors \cite{Riess/etal:2011,Efstathiou:2013}. 

The highly significant tension  for the combined data set within the $\Lambda$CDM model however warrants further investigation: if we were to interpret the number in terms of Gaussian standard deviations it would correspond to 4.9 $\sigma$. Of course at this level it is very important to know  extremely well the tails of the distributions, here we have assumed that all errors on measurements of $H_0$ and $H(z)$  are Gaussian and this might not be exact.
The intermediate redshift measurements  from the cosmic chronometers project, seems not to be explicitly in tension either with the $z=0$ nor with the CMB.
 Elsewhere (Cuesta et al. 2014 in prep.) we will consider other probes of the expansion history that cover a similar redshift range: baryon acoustic oscillations and supernovae.   Moreover significant improvements in the $H(z)$ measurements is forthcoming (Moresco et al. 2014 in prep.), the significantly smaller error-bars might be able  to localise or to point to a possible reason for the tension. 

We have also explored how smooth the expansion history data are. This is motivated by the fact that we wanted to investigate if with  cosmology independent expansion history data we could see departures from a cosmological constant prediction. Using gaussian processes we concluded that the data are smooth with no sign of sharp deviations from a constant equation of state of dark energy over the redshift range $0.1\lesssim  z < 1.2$.    

\section*{Acknowledgements}

LV is supported by FP7- IDEAS Phys.LSS 240117. LV and RJ are supported by MICINN grant FPA2011-29678.

\section*{Appendix: Useful formulas}

For completeness we  report here the expressions for $H(z)$ for a homogeneous and isotropic FRW Universe.\\

For a flat universe, generic equation of state parameter for dark energy
\begin{equation}
H(z)=H_0 (1+z)^{3/2}\sqrt{\Omega_m+\Omega_{\Lambda}\exp\left[3\int_{0}^{z}\frac{w(z')}{(1+z')}dz'\right]}\,;
\end{equation}

for a non-flat Universe, equation of state parameter for dark energy given by the $w_0, w_a$ parameterization:
\begin{equation}
H(z)=H_0\left\{ \Omega_m (1+z)^3+\Omega_k(1+z)^2+\Omega_{\Lambda} (1+z)^{3(1+w_0+w_a})\exp[-3w_az/(1+z)]\right\}^{1/2}\,;
\end{equation}
of course for a flat $\Lambda$CDM model we have:
\begin{equation}
H(z)=H_0 \sqrt{\Omega_m(1+z)^3+(1-\Omega_m)}\,.
\end{equation}

\end{document}